\documentstyle[sprocl]{article}

\input{psfig.tex}

\def\Journal#1#2#3#4{{#1} {\bf #2}, #3 (#4)}

\def\NCA{\em Nuovo Cimento}

\def\NPB{{\em Nucl. Phys.} B}
\def\PLB{{\em Phys. Lett.}  B}
\def\PRL{\em Phys. Rev. Lett.}
\def\PRD{{\em Phys. Rev.} D}

\def\be{\begin{equation}}
\def\ee{\end{equation}}
\def\bea{\begin{eqnarray}}
\def\eea{\end{eqnarray}}

\newcommand{\preprint}[1]{\noindent\hfill\hbox{#1}\vskip 10pt}

\begin{document}
\preprint{Preprint Number: \parbox[t]{40mm}{ADP-98-033/T306}}

\title{Dimensionally Regularized Study of Nonperturbative\\
	Quenched QED }
\author{
   Andreas W.\ Schreiber, Tom Sizer and Anthony G.\ Williams
}

\address{
   Special Research Centre for the Subatomic Structure of Matter \\
	 and \\
   Department of Physics and Mathematical Physics, \\
   University of Adelaide, 5005, Australia \\
   E-mail: awilliam@physics.adelaide.edu.au
}

\maketitle\abstracts
{
  We study the dimensionally regularized fermion propagator  
  Dyson-Schwinger equation in quenched nonperturbative QED in an  
  arbitrary covariant gauge using the Curtis-Pennington vertex and  
  perform nonperturbative renormalization numerically.  The  
  nonperturbative fermion propagator is solved in $D\equiv 4-2\epsilon
  <4$ dimensional Euclidean space for a large number of values of  
  $\epsilon$ for two values of the coupling, $\alpha = 0.6$ and  
  $\alpha=1.5$. Results for $D=4$ are then obtained by extrapolation to
  $\epsilon\to 0$.  We compare these results against previous studies
  employing a modified ultraviolet cut-off regularization and find
  agreement to within   the numerical precision of the present
  calculations.
}

\section{Introduction}
\label{sec_intro}

The Dyson-Schwinger equations (DSE) are an infinite tower of integral
equations that self-consistently relate the Green's functions of a
field theory.  They provide an alternative to lattice gauge theory
\cite{Rothe} for investigating nonperturbative phenomena such as
dynamical chiral symmetry breaking and dynamical mass generation. The
advantage of such a continuum formulation is that it is at times
possible to obtain analytical insights, there is no limit to the
momentum range that can be studied and one is able to compare
different regularization schemes.
The disadvantage is that the tower has to be truncated to those
equations involving Green's functions with relatively few external
legs: consequently some of the Green's functions appearing in
the remaining equations are no longer determined self-consistently by
the DSE's and we must make an educated guess (Ansatz) regarding their
form.

As an example, quantum electrodynamics (QED) in four dimensions in
the DSE formalism \cite{TheReview,MiranskReview,FGMS} admits a
dynamical mass above a certain critical coupling. The coupled
integral equations determining the photon and fermion propagators are
completely closed once the photon-fermion proper vertex is specified.
The quenched approximation, in which the photon propagator is bare and
the coupling does not run with momentum, reduces the study of these
coupled equations to that of the fermion propagator: for
computational reasons it is almost universally used at present. The
vertex Ansatz is constrained by the need to respect the discrete
symmetries and the Ward-Takahashi identities \cite{WTI} (WTI), have no
artificial kinematic singularities, satisfy the requirements of
perturbative
multiplicative renormalizability (MR), and agree with perturbation
theory in the weak coupling limit. Furthermore the gauge dependence
of the resulting fermion propagator should be consistent with the
Landau-Khalatnikov transformation \cite{LKTF}. Ultimately, observables
like the critical coupling should be verified to be
gauge-independent, once the effect of the regulator is removed.

A number of discussions of the choice of proper vertex in quenched
QED can be found in the literature, e.g.,
Refs.~\cite{BC,CPI,CPII,CPIII,CPIV,dongroberts,BP1,BP2,Kiz_et_al}.
Here we concentrate on the Curtis-Pennington (CP) vertex, which
satisfies both the WTI and the constraints of multiplicative
renormalizability.  With a bare vertex (which breaks both gauge
invariance and MR) the critical coupling of quenched QED differs by
approximately 50\% when calculated in the Feynman and Landau gauge.
This should be compared to a difference of less than 2\% if this
quantity is calculated with the Curtis-Pennington vertex
\cite{CPIV,ABGPR}.  Similar results are obtained in \cite{qed4_hrw}.
These calculations used a form of the ultraviolet (UV) regulated DSE
with a gauge covariance correction, where an obvious gauge covariance
violating term arising from the cut-off was omitted.

Clearly, the gauge dependence of the critical coupling is decreased,
but not eliminated, through the use of a photon-fermion proper vertex
which satisfies the WTI. The question arises whether the
remaining slight gauge dependence in the critical coupling is
primarily due to limitations of the vertex itself or whether it is
due to the use of a UV cut-off regulator in these calculations.
Bashir and Pennington \cite{BP1,BP2} have pursued the first of these
alternatives and have obtained, within a cut-off regularized theory,
further restrictions on the transverse part of the vertex which
ensure by construction that the critical coupling indeed becomes
strictly gauge independent. It is not clear that adjusting the vertex
to remove an unwanted gauge-dependence is the most appropriate
procedure when using a gauge-invariance (and  Poincar\'e-invariance)
violating regularization scheme.

We explore the second alternative by studying renormalized
quenched QED numerically using the CP vertex and dimensional
regularization \cite{tHooft-Velt} where gauge invariance is
explicitly maintained. To our knowledge this is the first complete
nonperturbative demonstration of dimensional regularization and
renormalization, and follows procedures first developed using the DSE
for a UV cutoff regulator in
Refs.~\cite{qed4_hw,qed4_hrw,qed4_hsw}.
The resulting nonlinear integral
equation for the fermion propagator is solved numerically in $D = 4 -
2 \epsilon  < 4$ Euclidean dimensional space.  Successive 
calculations with decreasing $\epsilon$ are then extrapolated to
$\epsilon=0$ and compared with the corresponding (modified and
unmodified) UV cutoff results.


\section{Formalism}
\label{sec_formalism}

We summarize the implementation of nonperturbative renormalization
using dimensional regularization within the context of
numerical DSE studies, adopting a notation similar to that used in
Refs.~\cite{qed4_hw,qed4_hrw,qed4_hsw}.  The formalism is presented in
Minkowski space and the Wick rotation into Euclidean space can then
be performed once the equations to be solved have been written down.
Note we use dimensional regularization in conjunction with off-shell
renormalization, rather than using the $MS$ or $\overline{MS}$
renormalization schemes, which can only be defined in a perturbative
context.

The renormalized inverse fermion propagator is defined through
\begin{eqnarray}                        \label{fermprop_formal}
            S^{-1}(\mu;p)  =   A(\mu;p^2) \not\!p - B(\mu;p^2)
            & = & Z_2(\mu,\epsilon) [\not\!p - m_0(\epsilon)]
                              - \Sigma'(\mu,\epsilon; p) \nonumber\\
            & = & \not\!p - m(\mu) - \widetilde{\Sigma}(\mu;p)\;\;\;,
\end{eqnarray}
where $\mu$ is the chosen renormalization scale, $m(\mu)$ is the value of the
renormalized mass at $p^2 = \mu^2$, $m_0(\epsilon)$ is the bare mass 
and $Z_2(\mu,\epsilon)$ is the 
wavefunction renormalization constant.  Due to the WTI for the fermion-photon
proper vertex, we have for the vertex renormalization constant
$Z_1(\mu,\epsilon)=Z_2(\mu,\epsilon)$. The renormalized and unrenormalized
fermion self-energies are denoted as $\widetilde{\Sigma}(\mu;p)$
and $\Sigma'(\mu,\epsilon;p)$ respectively.  These can be expressed in terms
of Dirac and scalar pieces, where for example
\begin{equation}
  \Sigma'(\mu,\epsilon; p) = \Sigma'_d(\mu,\epsilon; p^2) \not\!p
		     + \Sigma'_s(\mu,\epsilon; p^2)\;\;\;,
  \label{decompose}
\end{equation}
and similarly for $\widetilde{\Sigma}(\mu;p)$.
In the interests of notational brevity we do not explicitly indicate
the dependence on $\epsilon$ of the renormalized quantities
$A(\mu;p^2)$, $B(\mu;p^2)$ and $\widetilde{\Sigma}(\mu;p)$,
since for these and other renormalized quantities we are always interested
in their $\epsilon\to 0$ limit.  The renormalized mass function 
$M(p^2) \equiv B(\mu;p^2)/A(\mu;p^2)$ is renormalization point independent,
which follows straightforwardly from multiplicative renormalizability
\cite{qed4_hrw}. 

The renormalization point boundary condition
\begin{equation}
  \left. S^{-1}(\mu;p) \right|_{p^2 = \mu^2}
  = \not\!p - m(\mu)\:
\label{ren_point_BC}
\end{equation}
implies that $A(\mu;\mu^2) \equiv 1$ and $m(\mu) \equiv M(\mu^2)$ and yields the 
following relations between renormalized and unrenormalized self-energies
\begin{equation}\label{ren_BC}
  \widetilde{\Sigma}_{d,s}(\mu; p^2) =
    \Sigma'_{d,s}(\mu,\epsilon; p^2) - \Sigma'_{d,s}(\mu,\epsilon; \mu^2) 
     \;\;\;.
\end{equation}
Also, the wavefunction renormalization is given by
\begin{equation}
  Z_2(\mu,\epsilon) = 1 + \Sigma'_d(\mu,\epsilon; \mu^2)
\label{eq_Z2}
\end{equation}
and the bare mass $m_0(\epsilon)$ is linked to the renormalized mass 
$m(\mu)$ through
\begin{equation}
  m_0(\epsilon) = \left[ m(\mu) - \Sigma'_s(\mu,\epsilon; \mu^2) \right]
	/ Z_2(\mu,\epsilon)\;\;\;.
\label{baremass}
\end{equation}
It also follows from MR that under a renormalization point transformation
$\mu \to \mu^\prime$, $m(\mu^\prime) = M({\mu^\prime}^2)$ and
$Z_2(\mu^\prime,\epsilon) = A(\mu^\prime; \mu^2) \, Z_2(\mu,\epsilon)$
as discussed in \cite{qed4_hrw}.

The unrenormalized self-energy is given by the integral
\begin{equation} \label{reg_Sigma}
  \Sigma'(\mu,\epsilon; p) = i Z_1(\mu,\epsilon) [e(\mu) \nu^\epsilon]^2 \int
    \frac{d^Dk}{(2\pi)^D} \gamma^\lambda {S}(\mu;k)
      {\Gamma}^\nu(\mu; k,p)
      {D}_{\lambda \nu}(\mu;p-k)\:,
\end{equation}
where $\nu$ is an arbitrary mass scale introduced in $D$
dimensions  so that the renormalized
coupling $e(\mu)$ remains dimensionless.   Since we are here working in the
quenched approximation 
we have $Z_3(\mu,\epsilon)=1$, $e_0\equiv e(\mu)$,
and the renormalized  photon propagator ${D}^{\mu \nu}(\mu;q)$ is equal
to the bare photon propagator
\begin{equation}
  D^{\mu\nu}(q) = \left (
     -g^{\mu\nu} + \frac{q^\mu q^\nu}{q^2} 
     - \xi \frac{q^\mu q^\nu}{q^2} \right )
    \frac{1}{q^2}\:
\end{equation}
with $\xi$ being the covariant gauge parameter.  Finally, 
${\Gamma}^\nu(\mu;k,p)$
is the renormalized photon-fermion vertex for which we use the CP
Ansatz, namely ($q \equiv k - p$)
\begin{equation} \label{anyfullG_eqn}
  \Gamma^\mu(\mu;k,p) = \Gamma_{\rm BC}^\mu(\mu;k,p)
    + \tau_6(\mu;k^2,p^2,q^2) \left [\gamma^{\mu}(p^2-k^2)+(p+k)^{\mu}
{\not \! q}\right ]\:,
\end{equation}
where $\Gamma_{\rm BC}$ is the usual Ball-Chiu part of the vertex 
which saturates the
Ward-Takahashi identity \cite{BC}
\begin{eqnarray} \label{minBCvert_eqn}
  \Gamma^\mu_{\rm BC}(\mu;k,p) &=& \frac{1}{2}[A(\mu;k^2) +A(\mu;p^2)] 
\gamma^\mu
   \\
& &\hspace{-.2cm}+ \frac{(k+p)^\mu}{k^2-p^2}
      \left\{ [A(\mu;k^2) - A(\mu;p^2)] \frac{{\not\!k}+ {\not\!p}}{2}
	      - [B(\mu;k^2) - B(\mu;p^2)] \right\}
\nonumber
\end{eqnarray}
and the coefficient function $\tau_6$ is that chosen by Curtis and Pennington,
i.e.,
\begin{equation}
  \tau_6(\mu;k^2,p^2,q^2) = -\frac{1}{2}[A(\mu;k^2) - A(\mu;p^2)] / d(k,p)\:,
\label{CPgamma1}
\end{equation}
where
\begin{equation}
  d(k,p) = \frac{(k^2 - p^2)^2 + 
[M^2(k^2)  + M^2(p^2)  ]^2}{k^2+p^2}\:.
  \label{CPgamma2}
\end{equation}

The unrenormalized scalar and Dirac self--energies are extracted out of
the DSE, Eq.~(\ref{reg_Sigma}), by taking
$\frac{1}{4}{\rm Tr}$ of this equation, multiplied by 1 and 
$\not\!p/p^2$, respectively.  Note that we use the following
conventions \cite{Muta} for the Dirac algebra:
\begin{eqnarray*}
	g^\mu{}_\mu\> = D
	\hspace{1cm}&\Rightarrow&\hspace{1cm}
	\gamma^\mu \gamma_\mu = D, \quad
	\gamma^\mu \gamma^\nu \gamma_\mu = (2-D) \gamma^\nu \\
	{\rm Tr} [ {\bf 1} ] \> = \> 4 
	\hspace{1cm}&\Rightarrow&\hspace{1cm}
	{\rm Tr}\left [ \gamma^\mu \gamma^\nu \right ] = 4 g^{\mu \nu}.
\end{eqnarray*}

The integrands appearing in Eq.~(\ref{reg_Sigma}) only depend on the magnitude
of the internal fermion's momentum $k^2$ as well as the angle $\theta$ between
the fermion and photon momentum.  Hence the D-dimensional integrals
reduce to 2-dimensional ones, i.e.,

\begin{equation}
	\int d^D k \, f(k^2,p^2,k \cdot p) = \int d\Omega^{D-1} \times 
\int_0^\infty dk\,k^{D-1}\int_0^\pi d\theta \> \sin ^{D-2} \theta \>
 f(k^2,p^2,k \cdot p) 
\end{equation}
where 
\begin{equation}
	\int d\Omega^D = \frac{2\pi^{D/2}}{\Gamma(D/2)}
\end{equation}
is the surface area of a D-dimensional sphere.  Furthermore, it is
possible to express all the angular integrals in terms of a single
hypergeometric function so that it is only necessary to do the momentum
integral numerically, solving the renormalized fermion DSE by iteration.
Note that the momentum integration 
extends to infinity, necessitating a change in integration variables.  
The infinite range of the
integration also requires an extrapolation of $A(\mu;p^2)$ and $B(\mu;p^2)$ 
above the highest gridpoint.  We check insensitivity to this extrapolation by
comparing results obtained with a number of different extrapolation
prescriptions.  In addition, we use grids which extend some 20-30 orders of
magnitude beyond what is usually used in cut-off studies.  We have verified
the effect of the extrapolation to infinity is well-controlled.

\section{Results}
\label{sec_results}

We present solutions for the DSE for two values of the coupling
$\alpha = e_0^2/4\pi$, namely $\alpha=0.6$  and  $\alpha=1.5$, chosen
to correspond to couplings respectively well below and
above the critical coupling found in previous UV cut-off based studies.
Note that all results are quoted in terms of dimensionless
units, i.e., all mass and momentum scales can be simultaneoulsy multiplied
by any desired mass scale.

\begin{figure}[htb]
  \begin{minipage}[t]{5.5cm}
      \centering
      \psfig{figure=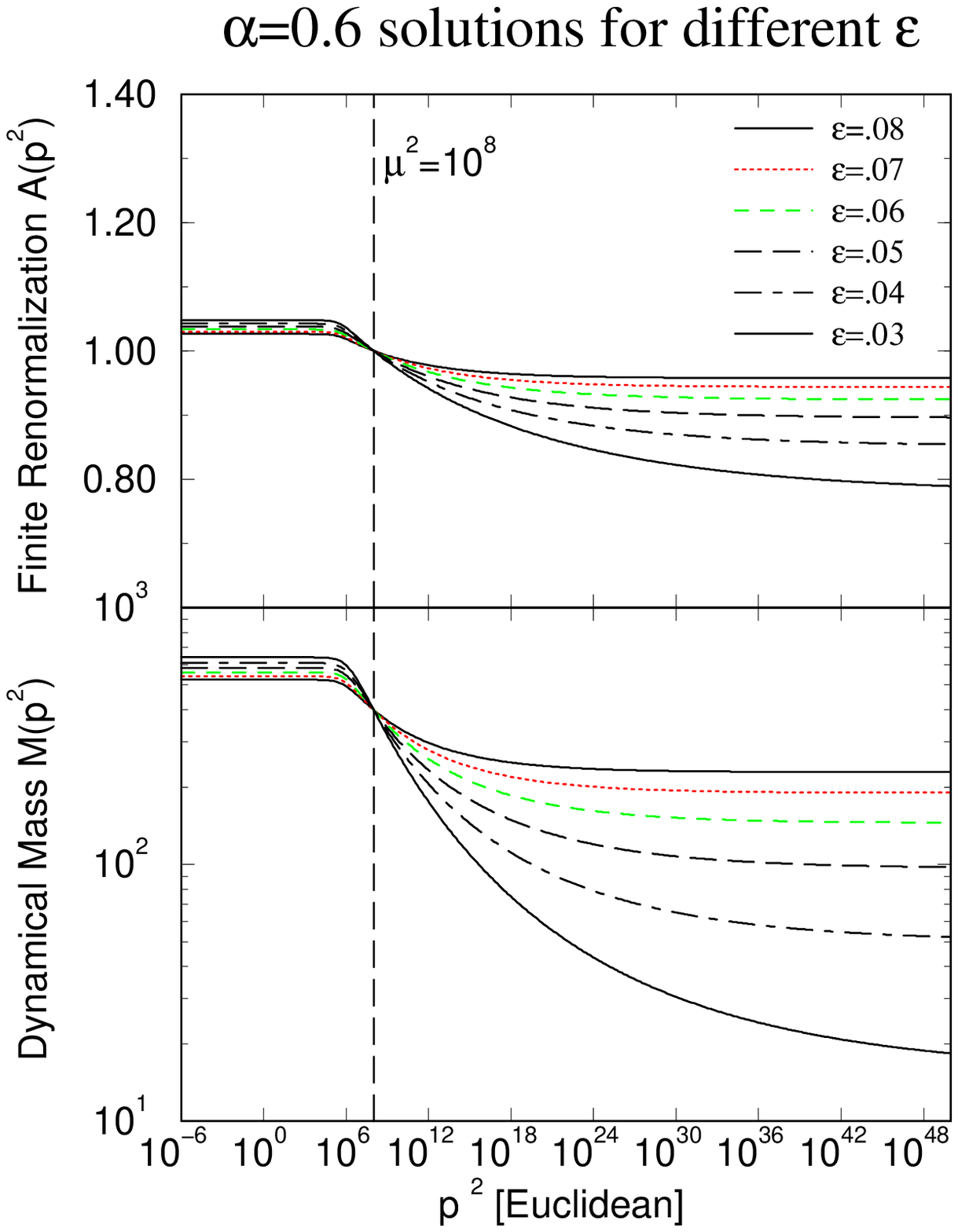,height=7.5cm}
      \begin{center}(a)\end{center}
  \end{minipage}
  \hfill\hfill
  \begin{minipage}[t]{5.5cm}
      \centering
      \psfig{figure=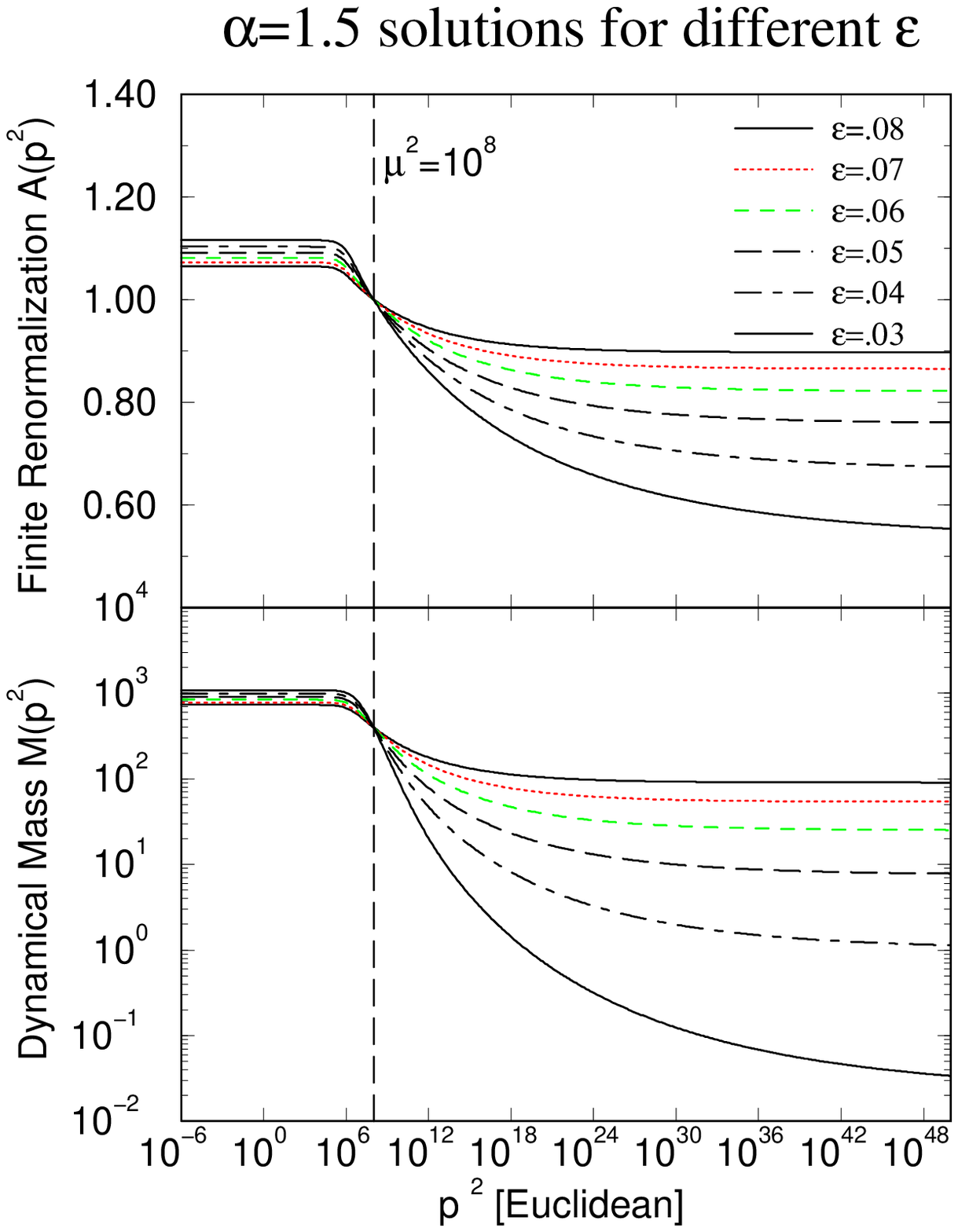,height=7.5cm}
      \begin{center}(b)\end{center}
  \end{minipage}
  \parbox{120mm}{\caption{
      The finite renormalization $A(p^2)$ and mass function $M(p^2)$
      for various choices of the regulator parameter $\epsilon$.
      Figure 1(a) has coupling $\alpha=0.6$ and 1(b) $\alpha=1.5$,
      with gauge parameter $\xi=0.25$, renormalization point $\mu^2=10^8$,
      renormalized mass $m(\mu)=400$ and scale $\nu=1$. In the low $p^2$
      region the smallest $\epsilon$ has the largest value of $M(p^2)$.
  \label{finite-eps-figs}}}
\end{figure}

Fig.~\ref{finite-eps-figs} show a family of solutions 
with the regulator parameter $\epsilon$ decreased from $0.08$ to $0.03$ for
the two values of the coupling.
We see that the mass function increases in strength in
the infrared and tails off faster in the ultraviolet as $\epsilon$ is reduced
or $\alpha$ is increased. 
Furthermore, it is important to note the strong
dependence on $\epsilon$, even though this parameter is already rather small.
As one would expect, the ultraviolet is most sensitive to this regulator,
however even in the infrared there is considerable dependence due to the
intrinsic coupling between these regions by the renormalization procedure.
This strong dependence on $\epsilon$ should be contrasted with the situation
in cut-off based studies where it was observed that already at rather modest
cut-offs ($\Lambda^2 \approx 10^{10}$) the renormalized functions $A$
and $M$ had reached their asymptotic limits.

Currently it is not possible
to decrease $\epsilon$ significantly below the values shown in
Fig.~\ref{finite-eps-figs} due to numerical limitations.
In order to extract the values of $A$ and
$M$ in four dimensions we extrapolate to $\epsilon=0$, which introduces
an added uncertainty in the final result.  However it is possible to
estimate this uncertainty by using the fact that in the limit
$\epsilon \rightarrow 0$ the renormalized quantities should become
independent of the arbitrary scale $\nu$, 
introduced to keep the coupling $\alpha$ dimensionless in $D$ dimensions.


\begin{figure}[htb]
  \begin{minipage}[t]{5.5cm}
      \centering
      \psfig{figure=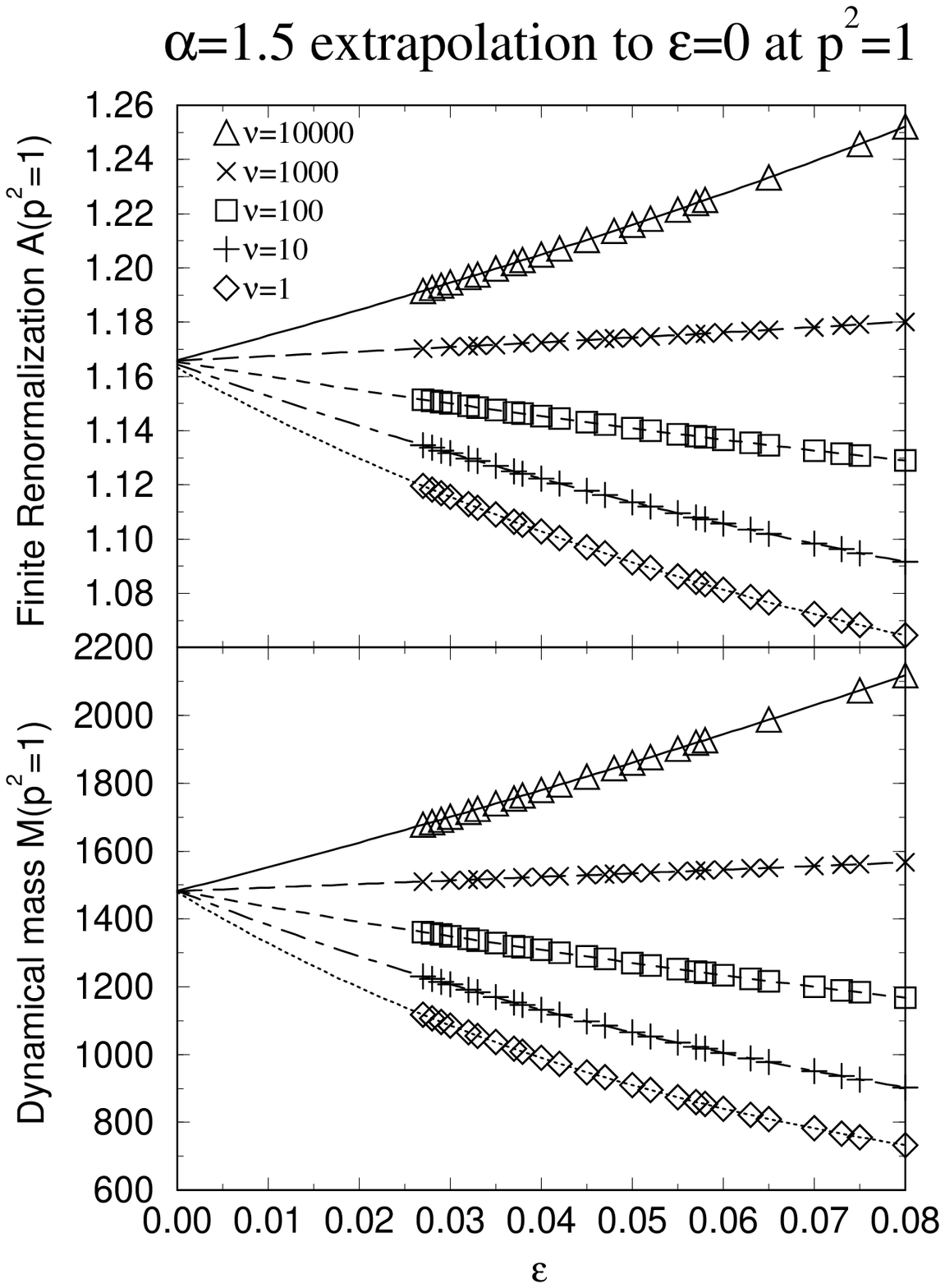,height=6.5cm}
      \begin{center}(a)\end{center}
  \end{minipage}
  \hfill
  \begin{minipage}[t]{5.5cm}
      \centering
      \psfig{figure=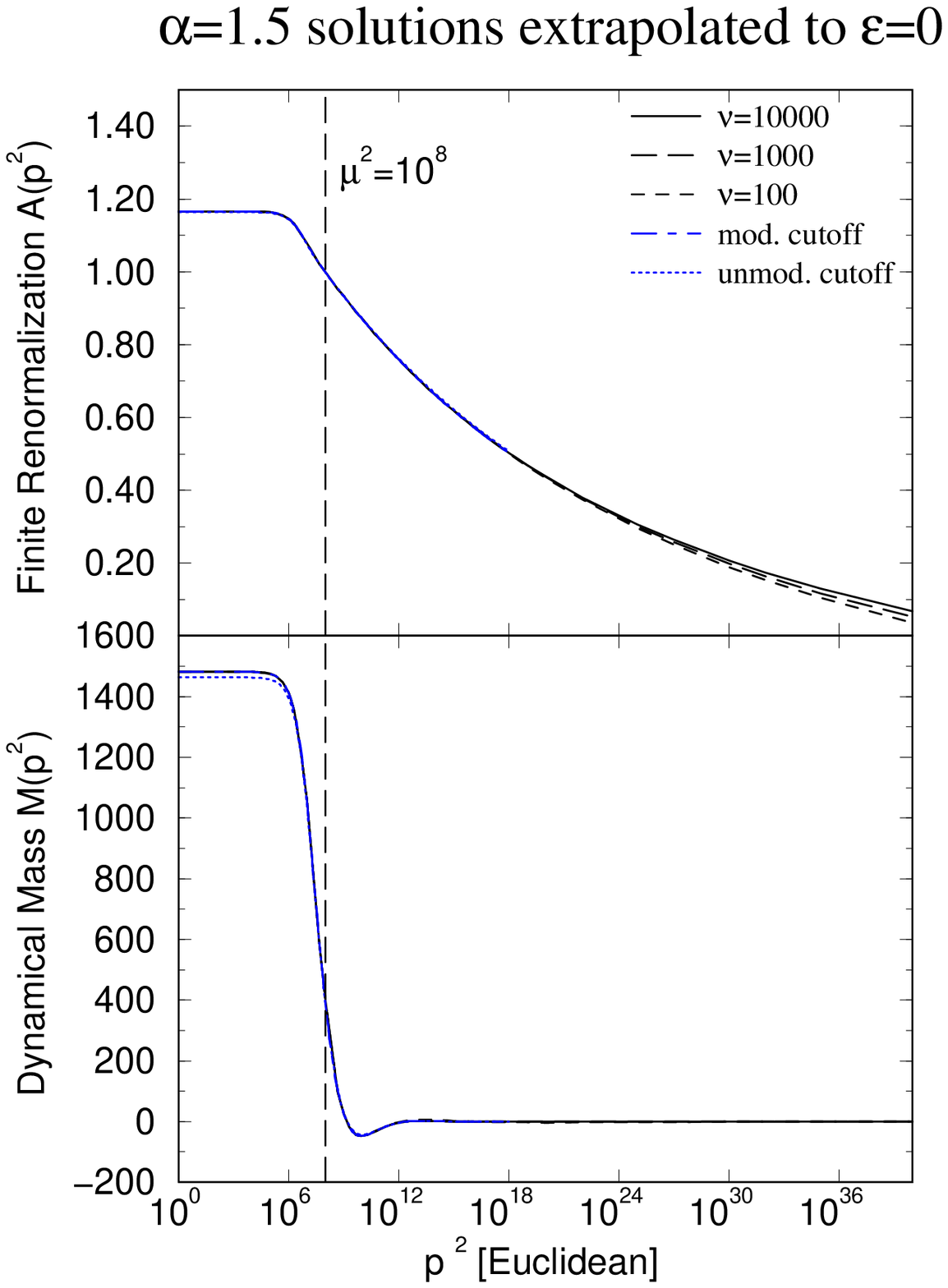,height=6.5cm}
      \begin{center}(b)\end{center}
  \end{minipage}
  \parbox{120mm}{\caption{
      (a) shows the finite renormalization $A(p^2)$ and mass
      function $M(p^2)$ evaluated at $p^2=1$ for various values
      of the regulator parameter
      $\epsilon$ and extrapolated to $\epsilon=0$ by fitting a polynomial
      cubic in $\epsilon$. Shown are five scales $\nu=1$, $10$,
      $100$, $1000$ and $10000$. All other parameters are those of Fig.~1.
      The different scales coincide at $\epsilon=0$ to an accuracy of
      approximately  $0.2\%$.
      (b) shows 
      the finite renormalization $A(p^2)$ and mass function $M(p^2)$
      extrapolated to $\epsilon=0$ at each momentum point for the three
      scales $\nu=100$, $1000$ and $10000$. Also shown are results
      obtained using a modified and unmodified UV cutoff. The modified UV
      cutoff and extrapolated results are indistinguishable in these plots,
      and differ from those of the unmodified UV cutoff in the IR.
  \label{eps0-fit}}}
\end{figure}

In Fig.~\ref{eps0-fit}a we show $A(p^2)$ and $M(p^2)$ evaluated
with $\alpha=e_0^2/4\pi=1.5$ in the infrared (at $p^2=1$) as a function of
$\epsilon$ for a range of values of $\nu$. The results at $\epsilon=0$
are extracted from cubic polynomial fits in $\epsilon$.  The
agreement between the different curves at $\epsilon=0$ is excellent,
being of the order of $0.2\%$.
In Fig.~\ref{eps0-fit}b we show the results for $\nu=100$, $1000$ and
$10000$ extrapolated to $\epsilon=0$ as a function of momentum
for $\alpha = 1.5$. Again the agreement is very good for a wide range of
$p^2$ and $\nu$.  Only in the ultraviolet region (above say $p^2 =
10^{12}$) are differences between the curves discernible.
The discrepancies for $\nu=1$ and $10$ cases were
somewhat greater in the UV, presumably reflecting greater non-linearity
in the fits for these cases that is manifest at $p^2 = 0$ in
Fig.~\ref{eps0-fit}a.
This is supported by the observation that these lower-scale cases
were more sensitive to changing the order of the polynomial the higher-scale
cases.
Also shown in Fig.~\ref{eps0-fit}b are the corresponding UV cutoff
results, both with and without a modification which used a
gauge-covariance fix to remove an obvious part of the gauge dependence
induced by the cutoff \cite{qed4_hrw}. The modified UV cutoff curve is
indistinguishable from the scaled dimensionally regularized ones,
while the unmodified
UV cutoff curve clearly deviates from the others in the IR. It should
also be noted that the oscillatory behaviour in the mass function first
noticed in \cite{qed4_hw} is reproduced in this work.

\section{Conclusions and Outlook}
\label{sec_conclusions}

We have reported the first detailed study of the numerical
renormalization of the fermion Dyson-Schwinger equation of QED
through the use of a dimensional regulator rather than a
gauge invariance-violating UV cut-off.  The initial results
are encouraging. Firstly, we have explicitly demonstrated
that the approach works and is independent of the intermediate
dimensional regularization scale as expected.  Secondly,
the previously obtained modified UV cut-off results are
reproduced with considerable precision.

A significant practical difference between the dimensionally and
UV cut-off regularized approaches is that in the former it is currently
necessary to perform an explicit extrapolation to $\epsilon=0$
whereas in the latter for a sufficiently large choice of UV cut-off
the results became independent of the cut-off.  We are presently
investigating whether it is numerically possible to extend
the studies to smaller values of $\epsilon$ in order to improve the
precision of the $\epsilon\to 0$ extrapolation.  In the meantime,
the extrapolation to $\epsilon=0$ appears to be well under control,
at least for values of the fermion momentum away from the ultraviolet
region.

Having demonstrated the numerical procedure of renormalization using
dimensional regularization, we plan to study chiral symmetry breaking
and in particular hope to extract the critical coupling as a function
of the gauge parameter, and to shed further light on the nature of
the chiral limit discussed in \cite{qed4_hsw}.  Results of this work
will be presented elsewhere.  The eventual aim is to extend this
treatment to the case of unquenched QED.

\section*{Acknowledgments}

This work was partially supported by grants from the Australian Research 
Council and by an Australian Research Fellowship.



\section*{References}
\begin{thebibliography}{99}
\bibitem{Rothe} H.J.\ Rothe,
    {\it Lattice Gauge Theories:  an Introduction\/},
    (World Scientific, Singapore, 1992).
\bibitem{TheReview}  C.D.~Roberts and A.G.~Williams,
    {\it Dyson-Schwinger Equations and their Application to Hadronic
    Physics\/}, in
    {\it Progress in Particle and Nuclear Physics, Vol.~33}
    (Pergamon Press, Oxford, 1994), p.~477.
\bibitem{MiranskReview} V.\ A.\ Miranskii,
    {\it Dynamical Symmetry Breaking in Quantum Field Theories},
    (World Scientific, Singapore, 1993).
\bibitem{FGMS} P.\ I.\ Fomin, V.\ P.\ Gusynin, V.\ A.\ Miransky and
    Yu.\ A.\ Sitenko, \Journal{\em Riv. Nuovo Cim.}{6}{1}{1983}.
\bibitem{WTI} J.C.~Ward, \Journal{\em Phys. Rev.}{78}{124}{1950};
    Y.~Takahashi, \Journal{\NCA}{6}{370}{1957}.
\bibitem{LKTF} L.~D.~ Landau and I.\ M.\ Khalatnikov,
    \Journal{\em Sov.\ Phys.\ JETP}{2}{69}{1956}
    [translation of \Journal{\em Zhur. Eksptl. i Teoret. Fiz.}{29}{89}{1955}];
    K.\ Johnson and B.\ Zumino, \Journal{\PRL}{3}{351}{1959}.
\bibitem{BC} J.S.~Ball and T.W.~Chiu, \Journal{\PRD}{22}{2542}{1980};
 \Journal{\PRD}{22}{2550}{1980}.
\bibitem{CPI} D.C.~Curtis and M.R.~Pennington, \Journal{\PRD}{42}{4165}{1990}.
\bibitem{CPII} D.C.~Curtis and M.R.~Pennington, \Journal{\PRD}{44}{536}{1991}.
\bibitem{CPIII} D.C.~Curtis and M.R.~Pennington, \Journal{\PRD}{46}{2663}{1992}.
\bibitem{CPIV} D.C.~Curtis and M.R.~Pennington, \Journal{\PRD}{48}{4933}{1993}.
\bibitem{dongroberts} Z.~Dong, H.~Munczek, and C.D.~Roberts,
     \Journal{\PLB}{333}{536}{1994}.
\bibitem{BP1} A.~Bashir and M.~R.~Pennington, \Journal{\PRD}{50}{7679}{1994}.
\bibitem{BP2} A.~Bashir and M.~R.~Pennington, \Journal{\PRD}{53}{4694}{1996}.
\bibitem{Kiz_et_al} A.\ Kizilers\"{u}, M.\ Reenders, and
    M.\ R.\ Pennington, \Journal{\PRD}{52}{1242}{1995}.
\bibitem{ABGPR} D.~Atkinson, J.C.R.~Bloch, V.P.~Gusynin, M.~R.~
    Pennington, and M.~Reenders, \Journal{\PLB}{329}{117}{1994}.
\bibitem{qed4_hw} F.T.\ Hawes and A.G.\ Williams, \Journal{\PRD}{51}{3081}{1995}.
\bibitem{qed4_hrw} F.T.\ Hawes, A.G.\ Williams and C.D.\ Roberts,
    \Journal{\PRD}{54}{5361}{1996}.
\bibitem{qed4_hsw} F.T.\ Hawes, T.\ Sizer and A.G.\ Williams,
    \Journal{\PRD}{55}{3866}{1997}.
\bibitem{tHooft-Velt} G. 't\ Hooft and M.J.G. Veltman, \Journal{\NPB}{44}{189}{1972}.
\bibitem{Muta} T.\ Muta, {\it Foundations of quantum chromodynamics
    -- An introduction to perturbative methods in gauge theories},
    (World Scientific, Singapore, 1987).

\end{thebibliography}
\end{document}